\documentclass[a4paper]{jpconf}
\usepackage{graphicx}
\usepackage{amssymb}
\usepackage{amsmath}
\usepackage{color}

\usepackage{mwe} 
\usepackage[caption=false]{subfig}

\newcommand{\Bb}{\textbf{B}}
\newcommand{\Eb}{\textbf{E}}
\newcommand{\sbb}{\textbf{s}}
\newcommand{\vb}{\textbf{v}}
\newcommand{\Pb}{\textbf{P}}

\begin{document}
\title{Simulation of Polarized Beams from Laser-Plasma Accelerators}

\author{Anna H\"utzen$^{1,2}$, Johannes Thomas$^{3}$, Andreas Lehrach$^{4,5}$, T.~ Peter Rakitzis$^{6,7}$, Alexander Pukhov$^{3}$, Liangliang Ji$^{8,9}$, Yitong Wu$^{8,10}$, Ralf Engels$^{5}$ and Markus B\"uscher$^{1,2}$}

\address{$^{1}$ Peter Gr\"unberg Institut (PGI-6), Forschungszentrum J\"ulich, Wilhelm-Johnen-Str.\,1, 52425 J\"ulich, Germany}

\address{$^{2}$ Institut f\"ur Laser- und Plasmaphysik, Heinrich-Heine-Universit\"at D\"usseldorf, Universit\"atsstr.\,1, 40225 D\"usseldorf, Germany}

\address{$^{3}$ Institut f\"ur Theoretische Physik I, Heinrich-Heine-Universit\"at D\"usseldorf, Universit\"atsstr.\,1, 40225 D\"usseldorf, Germany}

\address{$^{4}$ JARA-FAME (Forces and Matter Experiments), Forschungszentrum J\"ulich and RWTH Aachen University, 52056 Aachen, Germany}

\address{$^{5}$ Institut f\"ur Kernphysik (IKP), Forschungszentrum J\"ulich, Wilhelm-Johnen-Str.\,1, 52425 J\"ulich, Germany}

\address{$^{6}$ Department of Physics, University of Crete, 71003 Heraklion-Crete, Greece}

\address{$^{7}$ Institute of Electronic Structure and Laser, Foundation for Research and Technology-Hellas, 71110 Heraklion-Crete, Greece}

\address{$^{8}$ State Key Laboratory of High Field Laser Physics, Shanghai Institute of Optics and Fine Mechanics, Chinese Academy of Sciences, Shanghai 201800, China}

\address{$^{9}$ CAS Center for Excellence in Ultra-intense Laser Science, Shanghai 201800, China}

\address{$^{10}$ Center of Materials Science and Optoelectronics Engineering, University of Chinese Academy of Sciences, Beijing 100049, China}

\ead{a.huetzen@fz-juelich.de}

\begin{abstract}
 The generation of polarized particle beams still relies on conventional particle accelerators, which are typically very large in scale and budget. Concepts based on laser-driven wake-field acceleration have strongly been promoted during the last decades. Despite many advances in the understanding of fundamental physical phenomena, one largely unexplored issue is how the particle spins are influenced by the huge magnetic fields of plasma and, thus, how highly polarized beams can be produced. The realization of laser-plasma based accelerators for polarized beams is now being pursued as a joint effort of groups from Forschungszentrum Jülich (Germany), University of Crete (Greece), and SIOM Shanghai (China) within the ATHENA consortium. As a first step, we have theoretically investigated and identified the mechanisms that influence the beam polarization in laser-plasma accelerators. We then carried out a set of Particle-in-cell simulations on the acceleration of electrons and proton beams from gaseous and foil targets. We could show that intense polarized beams may be produced if pre-polarized gas targets of high density are employed. In these proceedings we further present that the polarization of protons in HT and HCl gas targets is largely conserved during laser wake-field acceleration, even if the proton energies enter the multi-GeV regime. Such polarized sources for electrons, protons, deuterons and $^{3}$He ions are now being built in Jülich. Proof-of-principle measurements at the (multi-)PW laser facilities PHELIX (GSI Darmstadt) and SULF (Shanghai) are in preparation.
\end{abstract}

\newpage

\section{Introduction}
\label{sect:Introduction}

In nuclear and particle physics, scattering experiments are widely used to study the structure and interaction of matter, and to test the standard model \cite{intro_1,intro_2}. In particular, the structure of sub-atomic particles like protons or neutrons can be explored by scattering experiments to get further insides of QCD \cite{intro_4}, to probe the nuclear structure \cite{intro_5}, or to investigate the dynamics of molecules \cite{intro_6, intro_7}. Spin-polarized particle beams are advantageous to get a better understanding of the nuclear structure and nuclear reactions \cite{intro_8}, to investigate symmetry violation, to interpret new asymmetries or to measure quantum numbers of new particles \cite{intro_3, intro_9, ILC_phys}. Up to now, in most cases high-energetic polarized beams are generated in conventional particle accelerators \cite{Mane2005}. In circular accelerators depolarizing spin resonances must be compensated by applying complex correction techniques maintaining the beam polarization \cite{RHIC, COSY, Huetzen2019}. Due to the very short interaction time between particle bunches and accelerating fields in linear accelerators, here depolarization effects are much less harmful compared to circular accelerators and storage rings.

Since conventional particle accelerators are typically very large in scale and budget, concepts based on laser-driven
wake-field acceleration from extremely intense laser pulses have strongly been promoted during the last decades. The ultimate goal
is to build highly compact and cost-effective laser-plasma accelerators (see Ref.\,\cite{EuPRAXIA}). 
However, one largely unexplored territory in this field is how particle spins are influenced by the huge magnetic fields inherently present in ultra-relativistic plasmas that are produced by Petawatt class laser systems interacting with dense gas or solid targets. Of special interest in this context is, which particular mechanisms could potentially lead to the production of highly polarized beams \cite{Huetzen2019, Buescher2019, Wu2019, Wen2019}. Generally speaking, one can distinguish between two extreme cases: i) either the magnetic field can align the spins of the accelerated particle beam; or ii) the spins are too inert, so that the plasma fields have no influence on the spin alignment. In the second case one would have to utilize a pre-polarized target whereas in the first case also an unpolarized target could be the right choice.

The effect of relevant physical processes on the degree of polarization of a particle beam has already been estimated analytically \cite{Thomas2019}. Here, in the classical and semi-classical limit, the acceleration of charged particles is treated within the framework of the classical field theory. The spin motion of single particle spins in a semi-classical limit is usually described by the Thomas-BMT equation \cite{Thomas, BMT}. This equation determines the spin precession around the local electromagnetic field lines. Other effects like the Stern-Gerlach force \cite{Stern-Gerlach} might become important for some artificial field configurations \cite{Flood2015}. Due to short time and length scales in our work, neither the Stern-Gerlach force nor the Sokolov-Ternov effect \cite{Sokolov-Ternov} are considered. Our results are obtained from the three-dimensional PIC code VLPL (Virtual Laser Plasma Lab) \cite{Pukhov2016,Vieira2016}. This code includes radiation reaction effects as well as the spin dynamics characterized by the Thomas-BMT equation.

In our three-dimensional simulations, we compare the temporal evolution of the polarization and the energy of pre-polarized protons in hydrogen-tritium (HT) and hydrogen-chloride (HCl) targets. In both cases the protons are accelerated by a high-intense Petawatt laser pulse creating an electron bubble-channel structure. Following the work of Baifei Shen et.\,al \cite{Shen2007}, where the acceleration of unpolarized protons to a maximum energy of $27$\,GeV is predicted for the HT case, we choose a similar simulation setup adding spin effects. Within the ATHENA (Accelerator Technology HElmholtz iNfrAstructure) consortium a dynamically pre-polarized ion source based on HCl gas is experimentally realized at Forschungszentrum J\"ulich \cite{Huetzen2019, Buescher2019, Rakitzis2004, Sofikitis2017}. Therefore, additional simulations for HCl are carried out in the frame of this work. The final experiments at the $10$\,PW laser system SULF at SIOM/Shanghai, aiming to observe a polarized particle beam from laser-generated HCl plasma, will be carried out  based on the obtained simulation results.

\section{Particle-in-cell Simulations Including Spin Dynamics}
\label{sect:PIC}
In our PIC simulations the single particle spins are treated in a semi-classical limit, where the Thomas-BMT equation 
\begin{align}
	\frac{d\sbb}{dt} = -\mathbf{\Omega}\times\sbb 
\label{PugaSpinNorm}
\end{align}
describes the precession of the spin $\sbb$ around the local electromagnetic fields $\Eb$ and $\Bb$. In cgs units the rotation frequency is simply \cite{Mane2005}
\begin{align}
	\mathbf{\Omega} = \frac{q}{mc}\left[\Omega_\mathrm{B}\Bb -\Omega_\mathrm{v}\left(\frac{\vb}{c}\cdot\Bb\right)\frac{\vb}{c} -\Omega_\mathrm{E}\frac{\vb}{c}\times\Eb\right],
\label{TBMT}
\end{align}	
where 
\begin{align}
	\Omega_\mathrm{B} = a+\frac{1}{\gamma}, && \Omega_\mathrm{v} = \frac{a \gamma}{\gamma+1}, && \Omega_\mathrm{E} = a +\frac{1}{1+\gamma} ,
\end{align}	
for a particle with rest mass $m$, charge $q$, velocity $\vb$, the gyromagnetic anomaly $a$, the energy $E=\gamma mc^{2}$, the Lorentz factor $\gamma$, and the speed of light $c$. The polarisation $\Pb$ of several particles is always calculated as 
\begin{align}
	\Pb = \frac{1}{N}\sum_{i=1}^{N} \sbb_\mathrm{i},
\end{align}
where $N$ could be the number of particles in a simulation cell within a certain energy range, or in another particular ensemble.

The basic idea for this work is inspired by a previous paper by Baifei Shen et.\,al \cite{Shen2007}, where three-dimensional PIC simulations show the trapping and acceleration of protons in an electron bubble-channel structure. The channel is driven by a highly intense, circularly polarized laser pulse with normalized peak amplitude $a_{\mathrm{0}}= eA/m_{\mathrm{e}}c =223$, which propagates through a near critical hydrogen-tritium plasma. In Ref.\,\cite{Shen2007} it has been shown that for a proper choice of laser and plasma parameters the light protons can be trapped by the laser and eventually be accelerated to energies far beyond 10\,GeV. In our PIC simulations we use a similar setup including the spin dynamics of pre-polarized protons in two simulation series. In the first simulation set we follow Ref.\,\cite{Shen2007} by choosing a hydrogen-tritium plasma with a hydrogen density of $n_{\mathrm{H}}=1\times 10^{20}$\,cm$^{-3}$ and a tritium density of $n_{\mathrm{T}}=1.4\times 10^{21}$\,cm$^{-3}$. The second series simulates a hydrogen-chloride gas with a hydrogen density of $n_{\mathrm{H}}=8.5\times 10^{19}$\,cm$^{-3}$ and an equal chloride density. In both simulations the electron density of $n_{\mathrm{e}}=1.5\times 10^{21}$\,cm$^{-3}$ is near-critical. To investigate whether an initial proton polarization is conserved, initially all protons are spin-aligned in \textit{y}-direction at the beginning of the simulation. The laser pulse is circularly polarized, has a wavelength of 800\,nm, a normalized laser amplitude of $a_{\mathrm{0}}=200$, a length of 10\,$\mu$m, and a focal spot size of 16\,$\mu$m. The simulation box is a co-moving frame of size 80\,$\mu$m $\times$ 80\,$\mu$m $\times$ 80\,$\mu$m, which is entered by the pulse from the left-hand side ($\xi=x-ct$-direction). There are 2500 $\times$ 100 $\times$ 100 cells in the simulation window and two PIC particles per species per cell. 
\newpage

\section{Results of Particle-in-cell Simulations}
\begin{figure}[b]
\centering
\subfloat[]{\label{t1_HT_density}\includegraphics[width=0.24\textwidth]{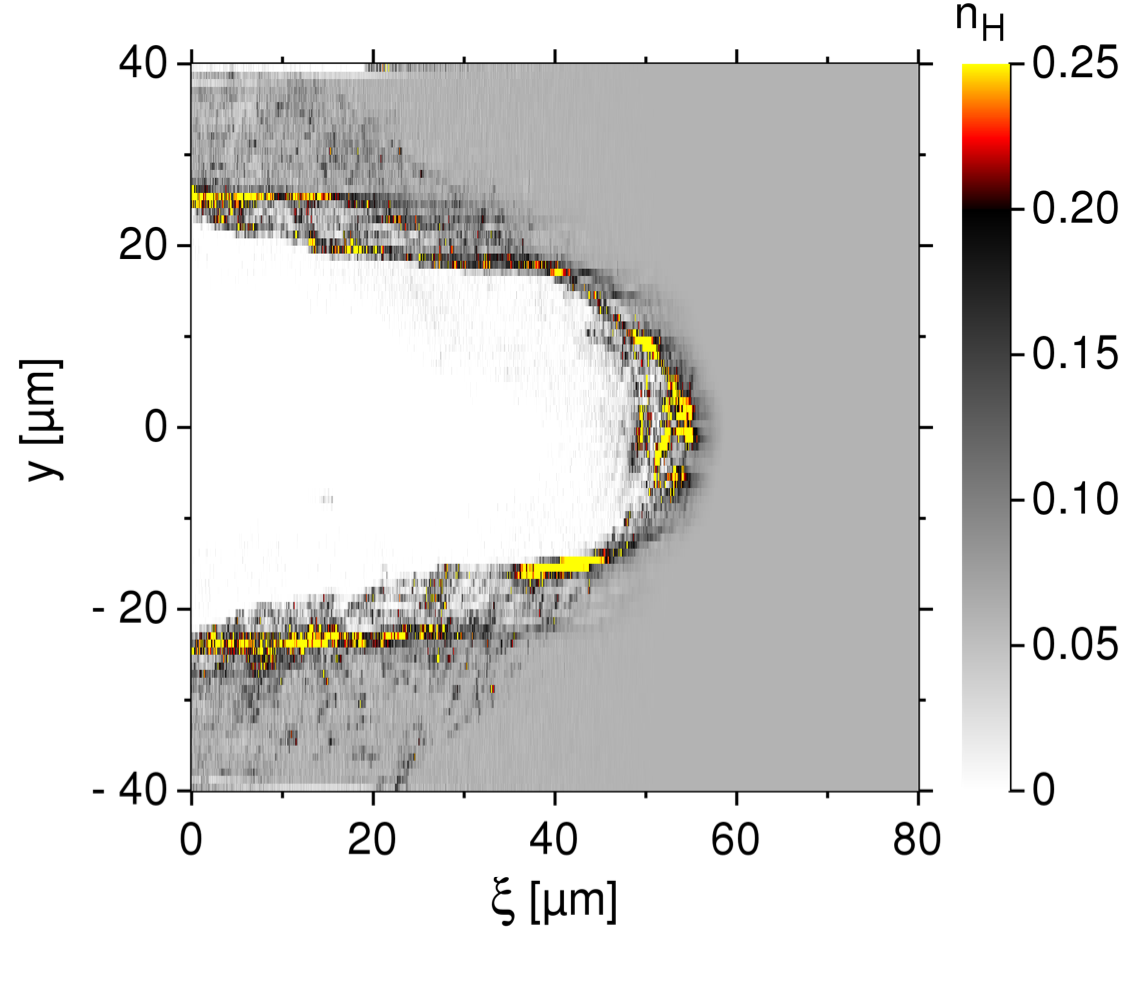}}
\hfill
\subfloat[]{\label{t1_HT_phase}\includegraphics[width=0.24\textwidth]{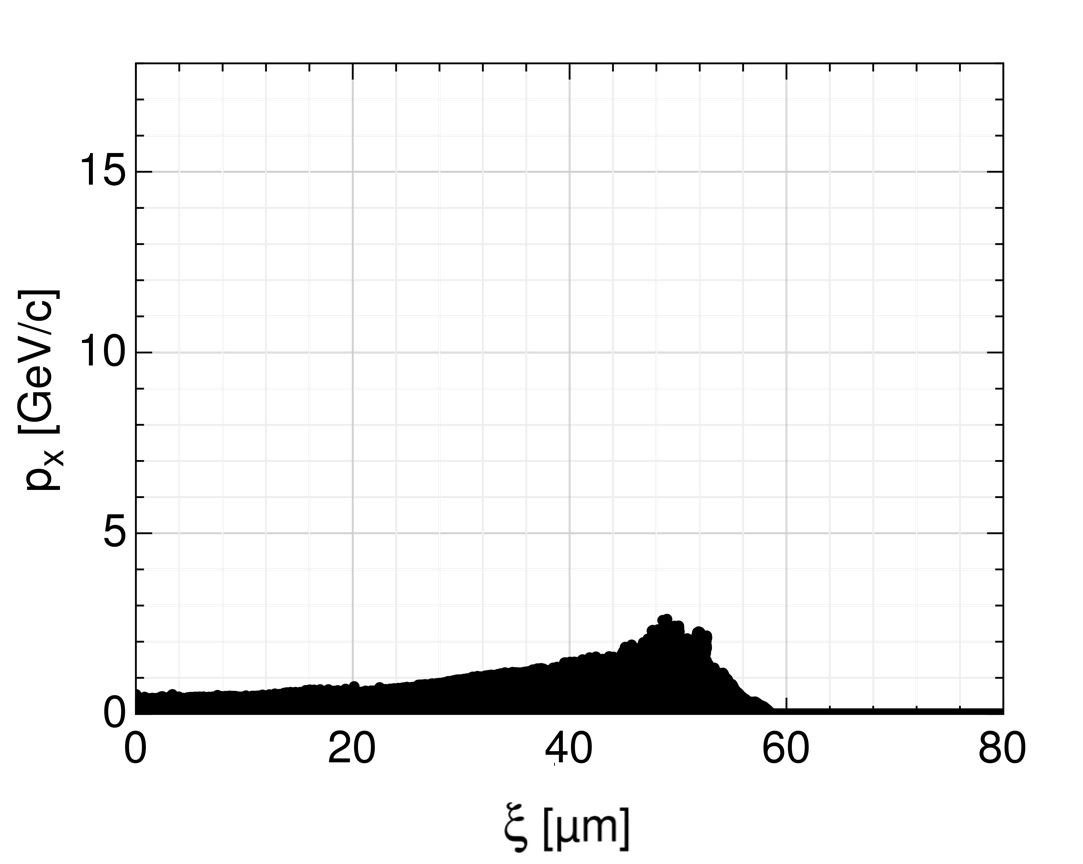}}
\hfill
\subfloat[]{\label{t1_HT_E_his}\includegraphics[width=0.24\textwidth]{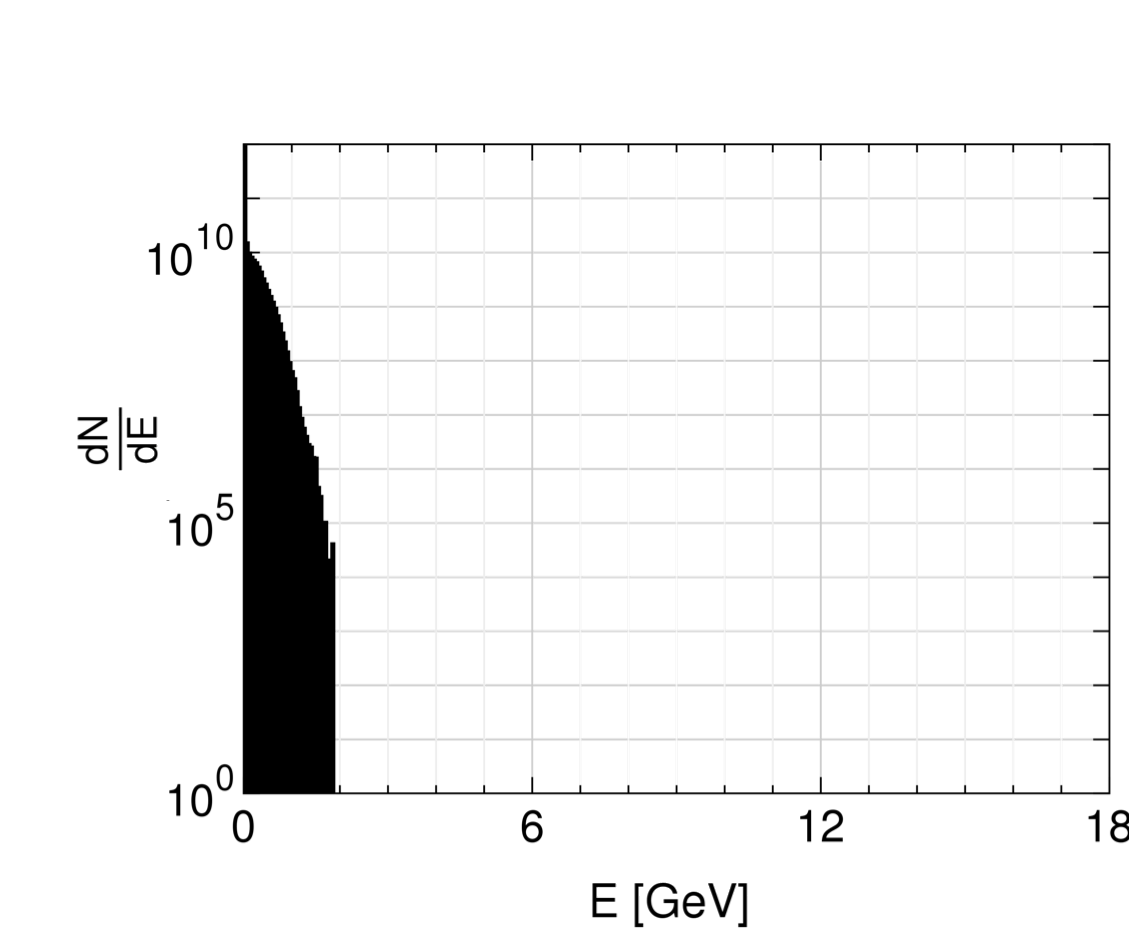}}
\hfill
\subfloat[]{\label{t1_HT_P_his}\includegraphics[width=0.24\textwidth]{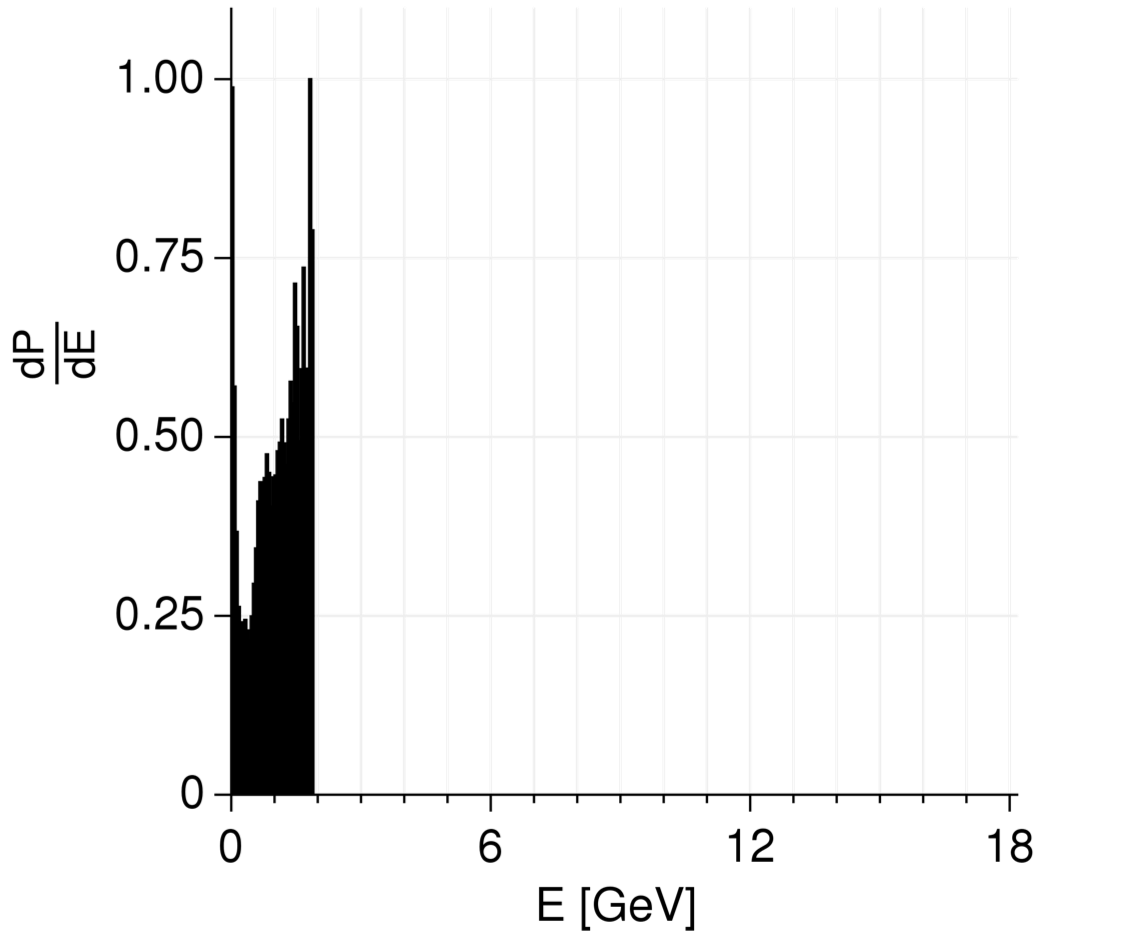}}\\
\subfloat[]{\label{t1_HCl_density}\includegraphics[width=0.24\textwidth]{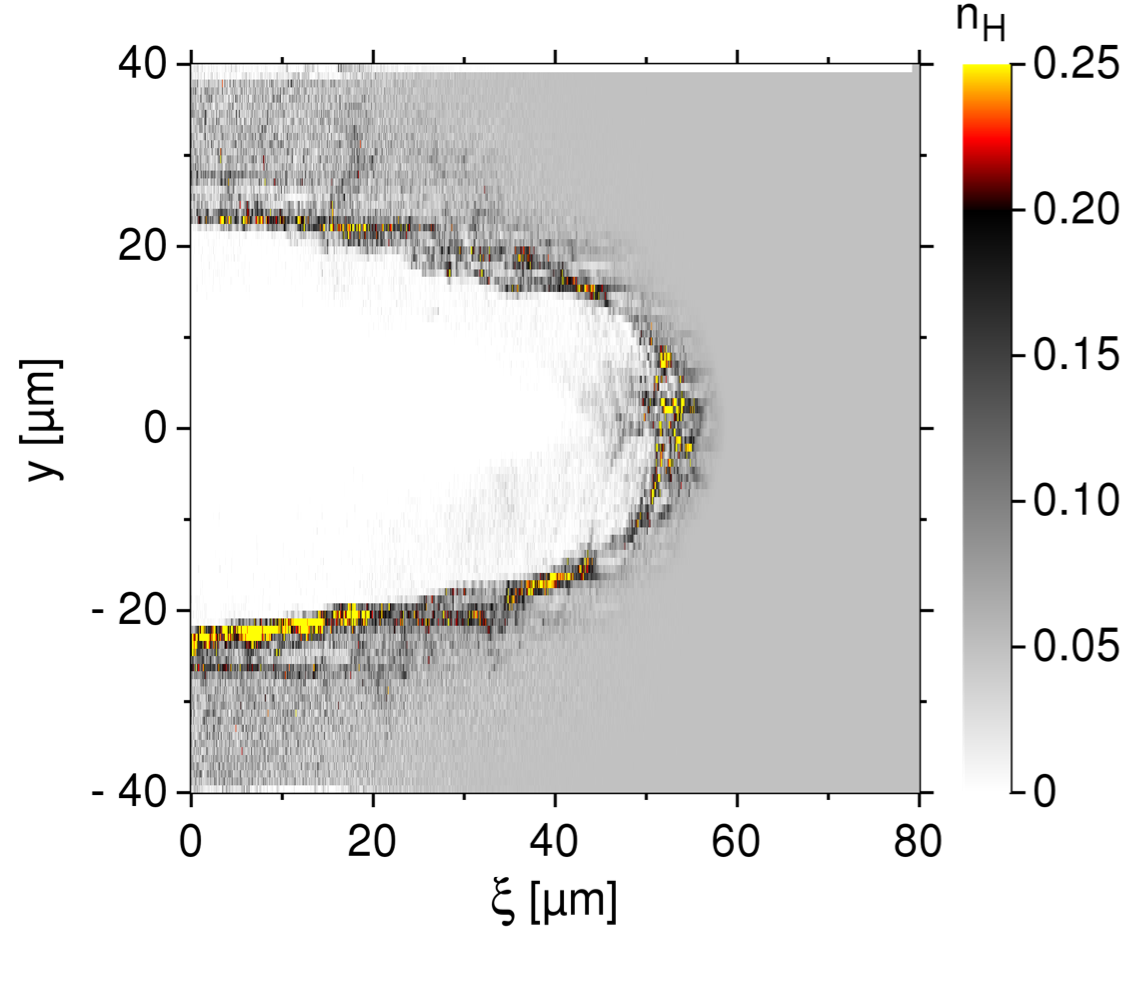}}
\hfill
\subfloat[]{\label{t1_HCl_phase}\includegraphics[width=0.24\textwidth]{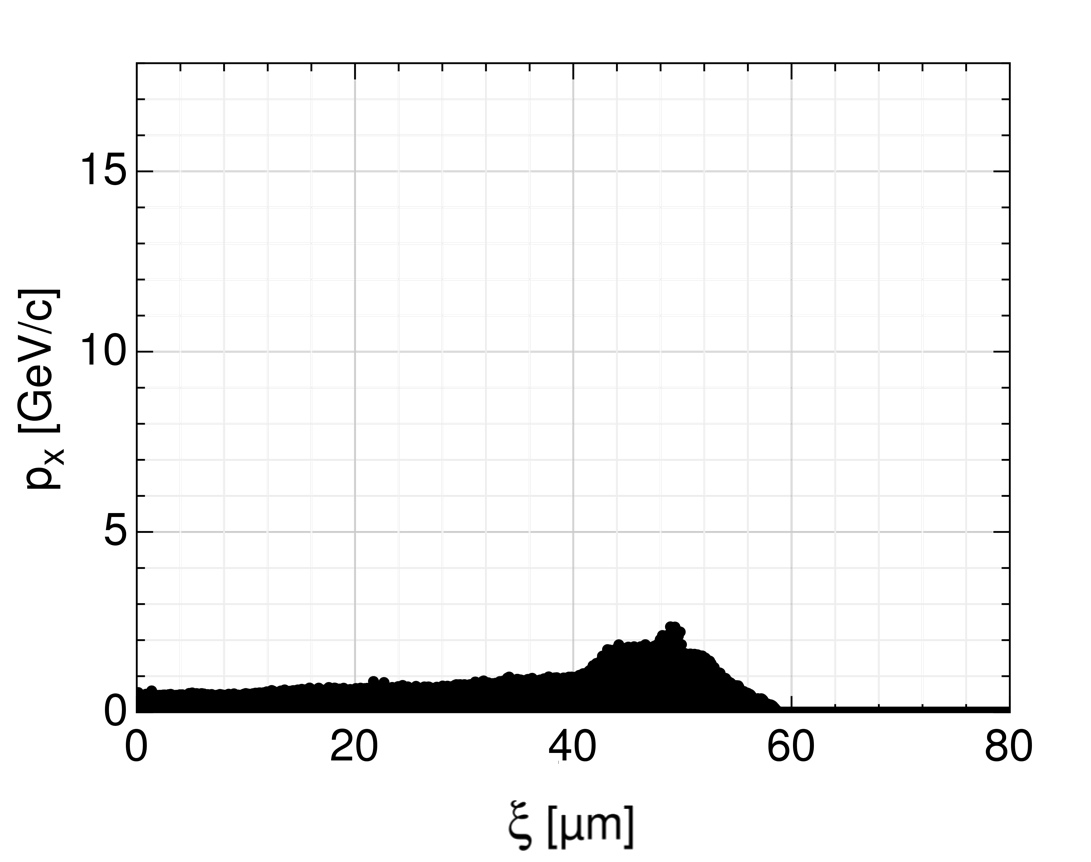}}
\hfill
\subfloat[]{\label{t1_HCl_E_his}\includegraphics[width=0.24\textwidth]{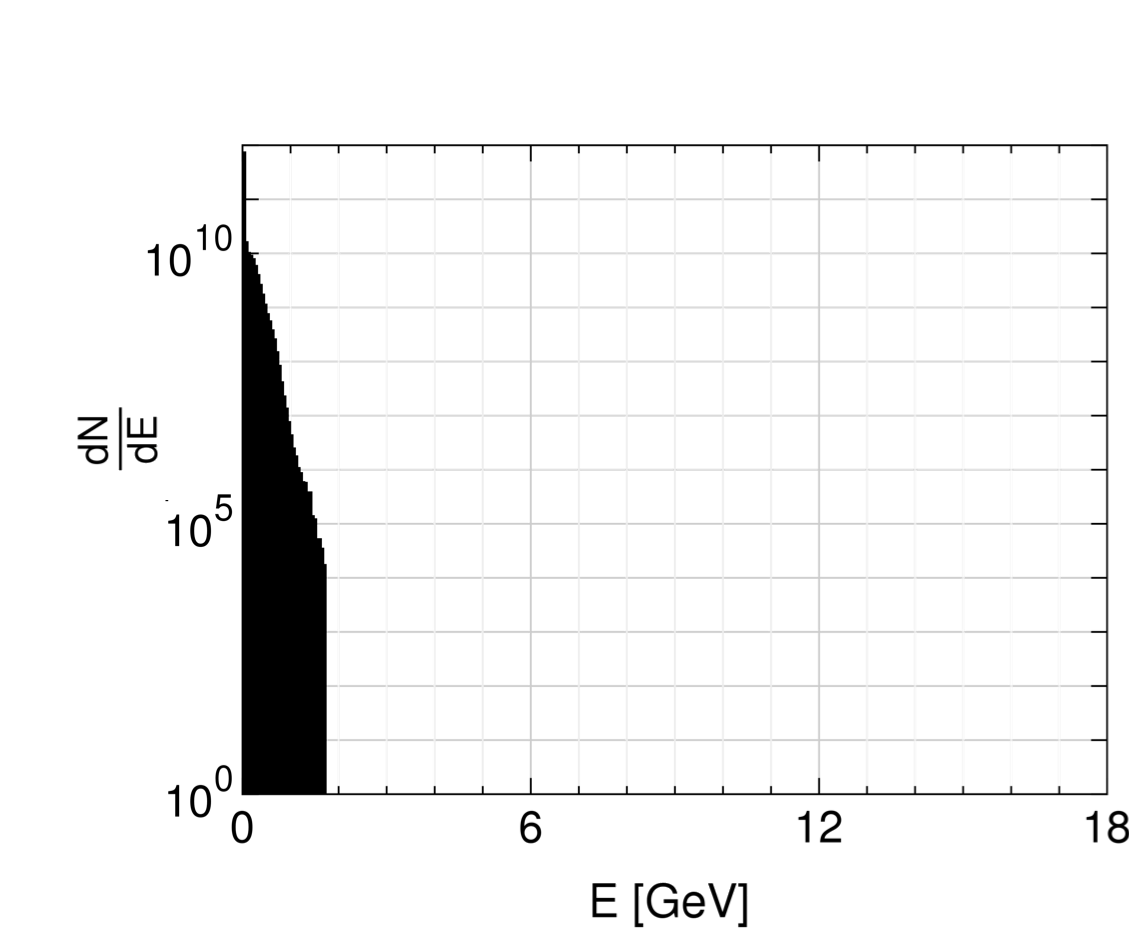}}
\hfill
\subfloat[]{\label{t1_HCl_P_his}\includegraphics[width=0.24\textwidth]{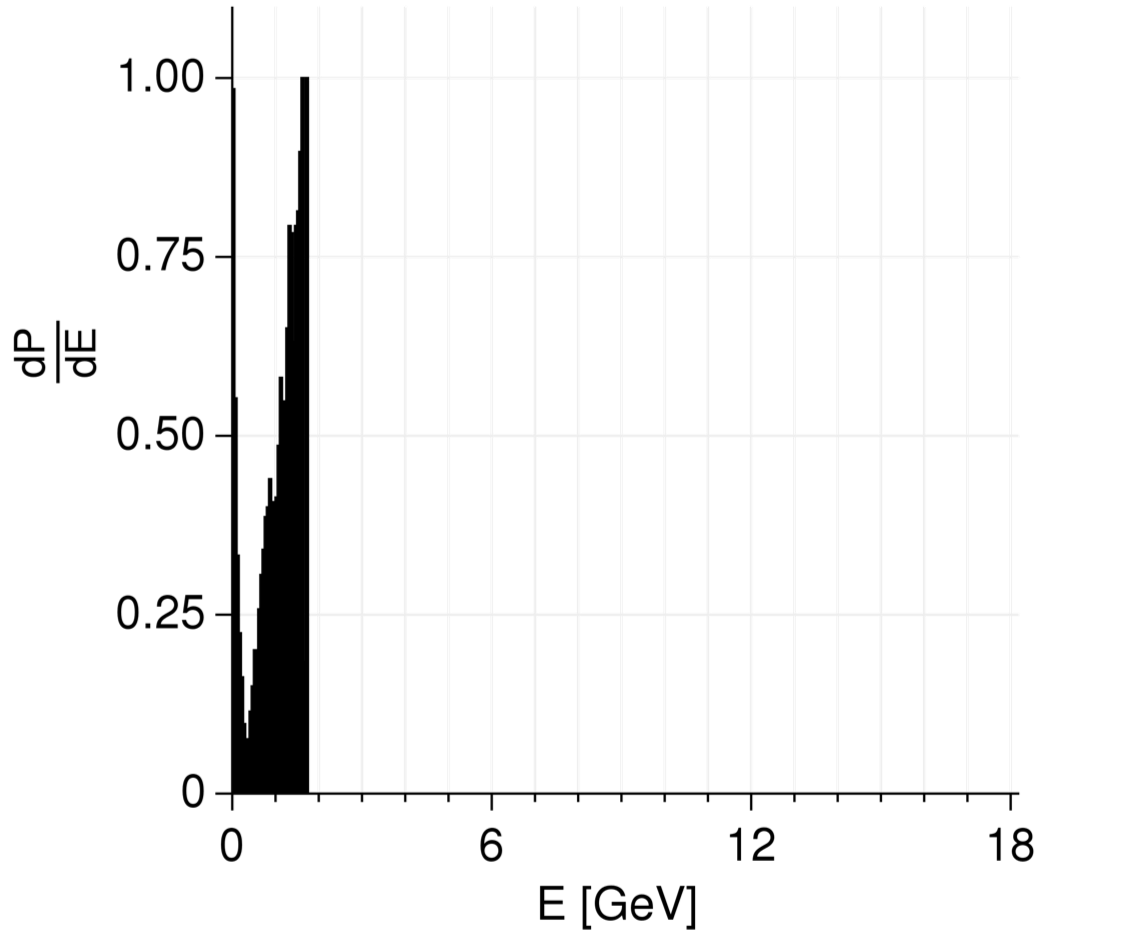}}
\caption{Simulation results for HT (upper plots) and HCl (lower) plasma at the time $t_{\mathrm{1}}=800$\,fs.
Figs.\,(\ref{t1_HT_density}),(\ref{t1_HCl_density}) show the proton densities, (\ref{t1_HT_phase}),(\ref{t1_HCl_phase}) the $p_\mathrm{x}$-$\xi$ phase-space, (\ref{t1_HT_E_his}),(\ref{t1_HCl_E_his}) the energy histograms and (\ref{t1_HT_P_his}),(\ref{t1_HCl_P_his}) the polarization distributions. The laser peak amplitude is located at $\xi=57.5$\,$\mu$m in both simulations.}
\label{Zeit1}
\end{figure}

\begin{figure}[t]
\centering
\subfloat[]{\label{t2_HT_density}\includegraphics[width=0.24\textwidth]{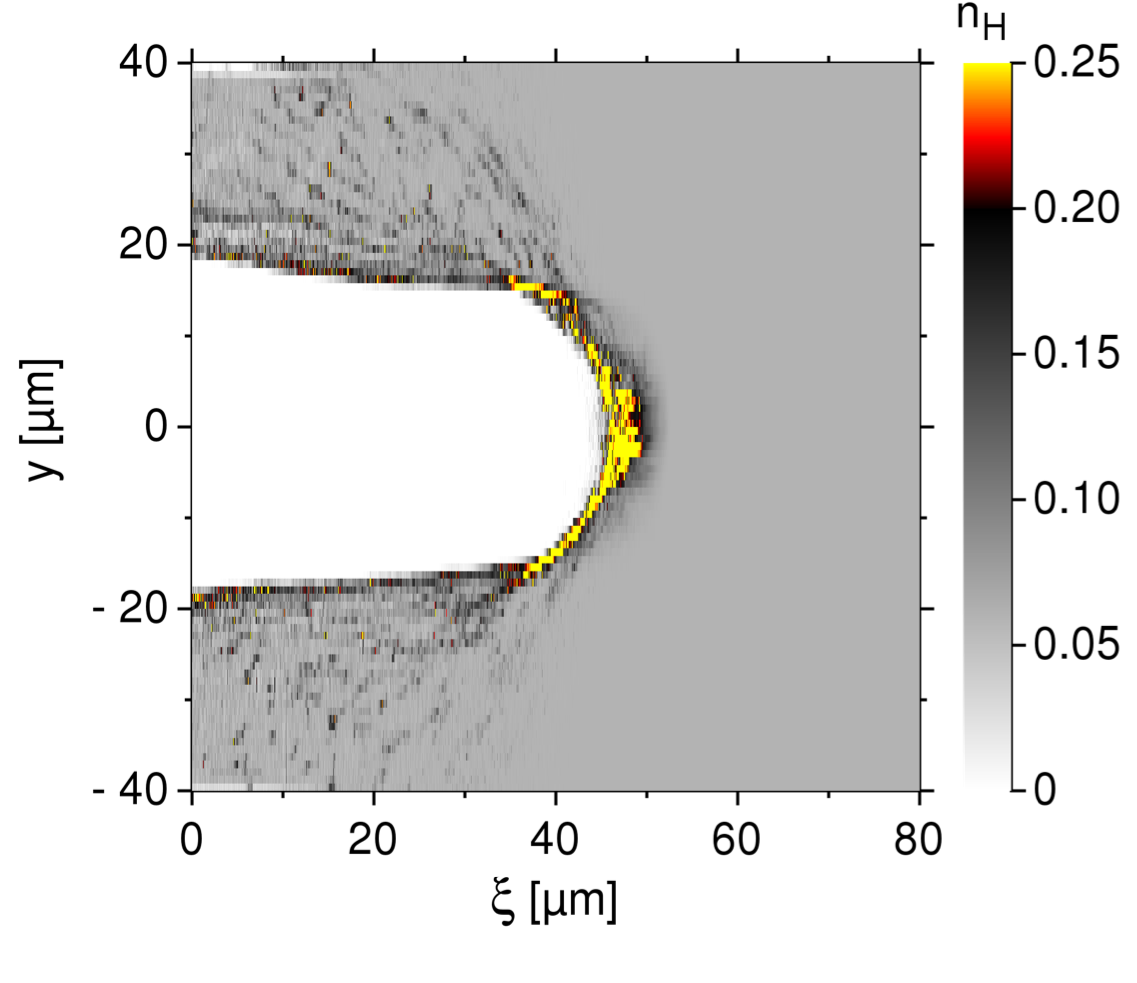}}
\hfill
\subfloat[]{\label{t2_HT_phase}\includegraphics[width=0.24\textwidth]{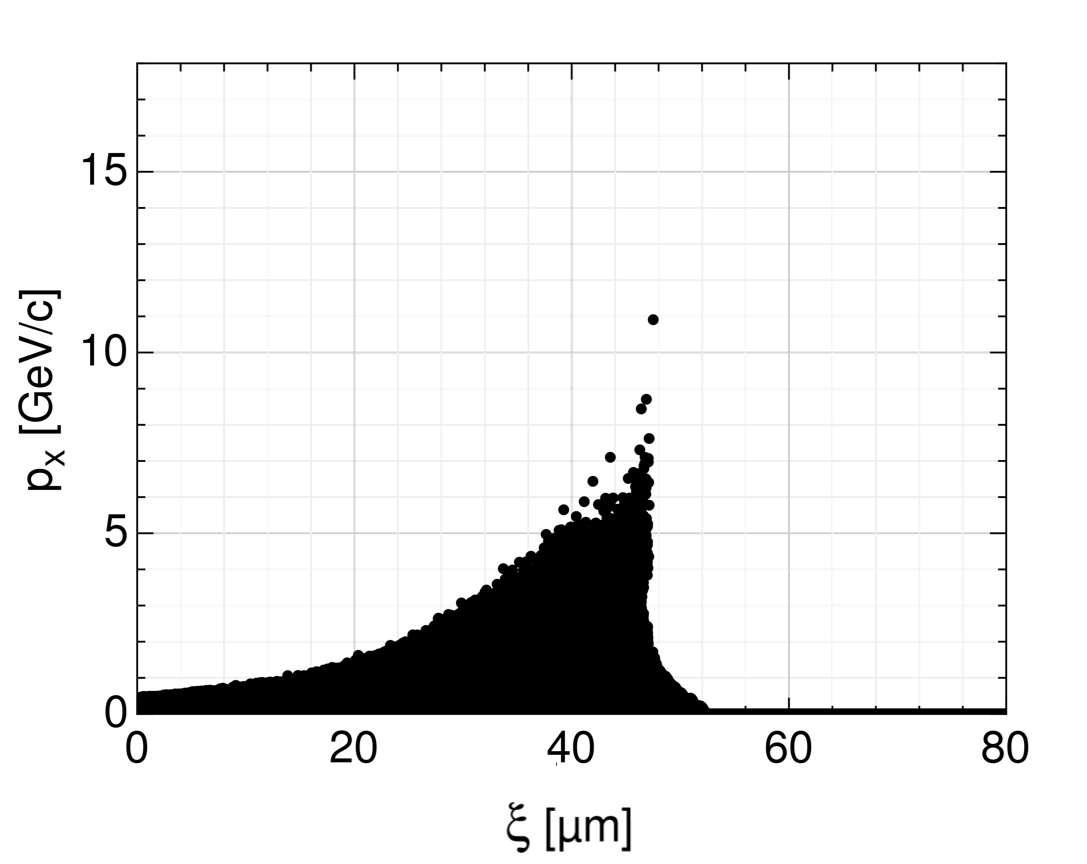}}
\hfill
\subfloat[]{\label{t2_HT_E_his}\includegraphics[width=0.24\textwidth]{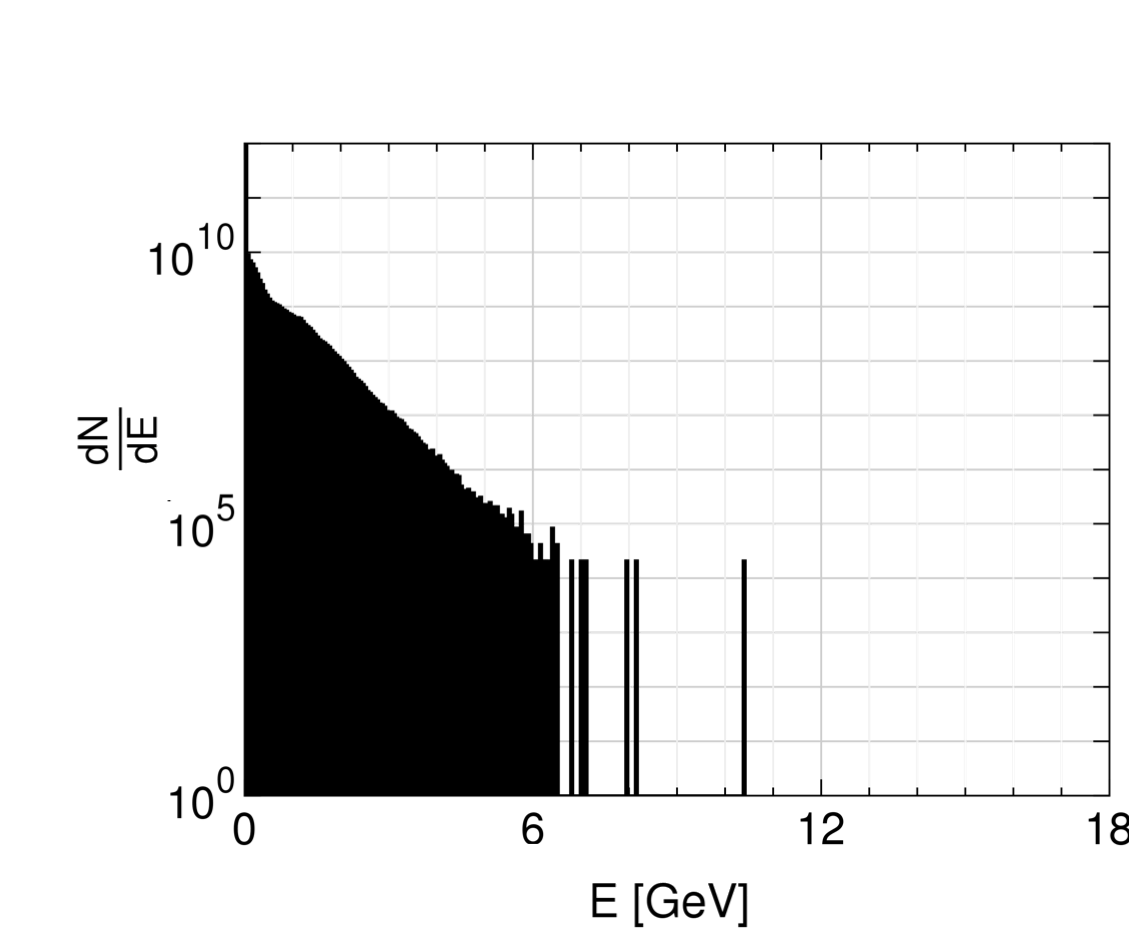}}
\hfill
\subfloat[]{\label{t2_HT_P_his}\includegraphics[width=0.24\textwidth]{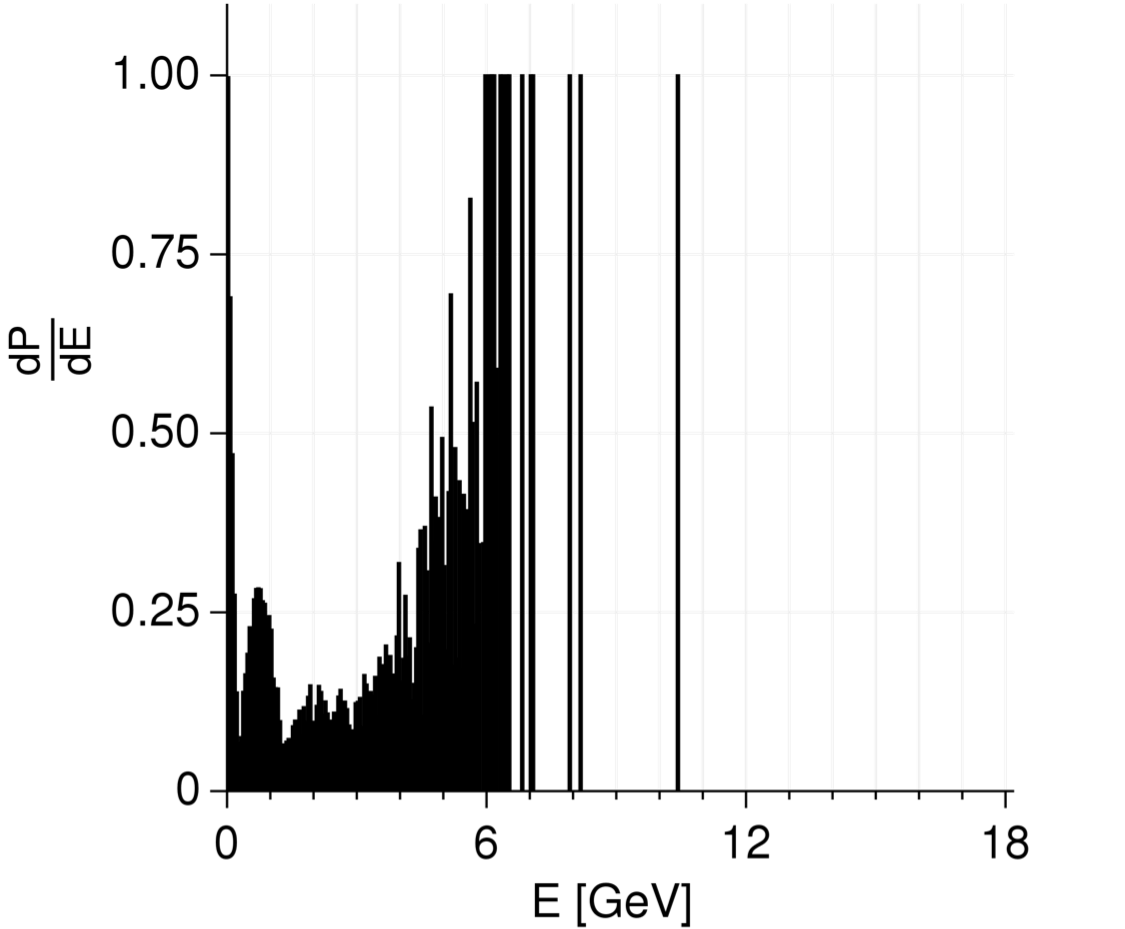}}\\
\subfloat[]{\label{t2_HCl_density}\includegraphics[width=0.24\textwidth]{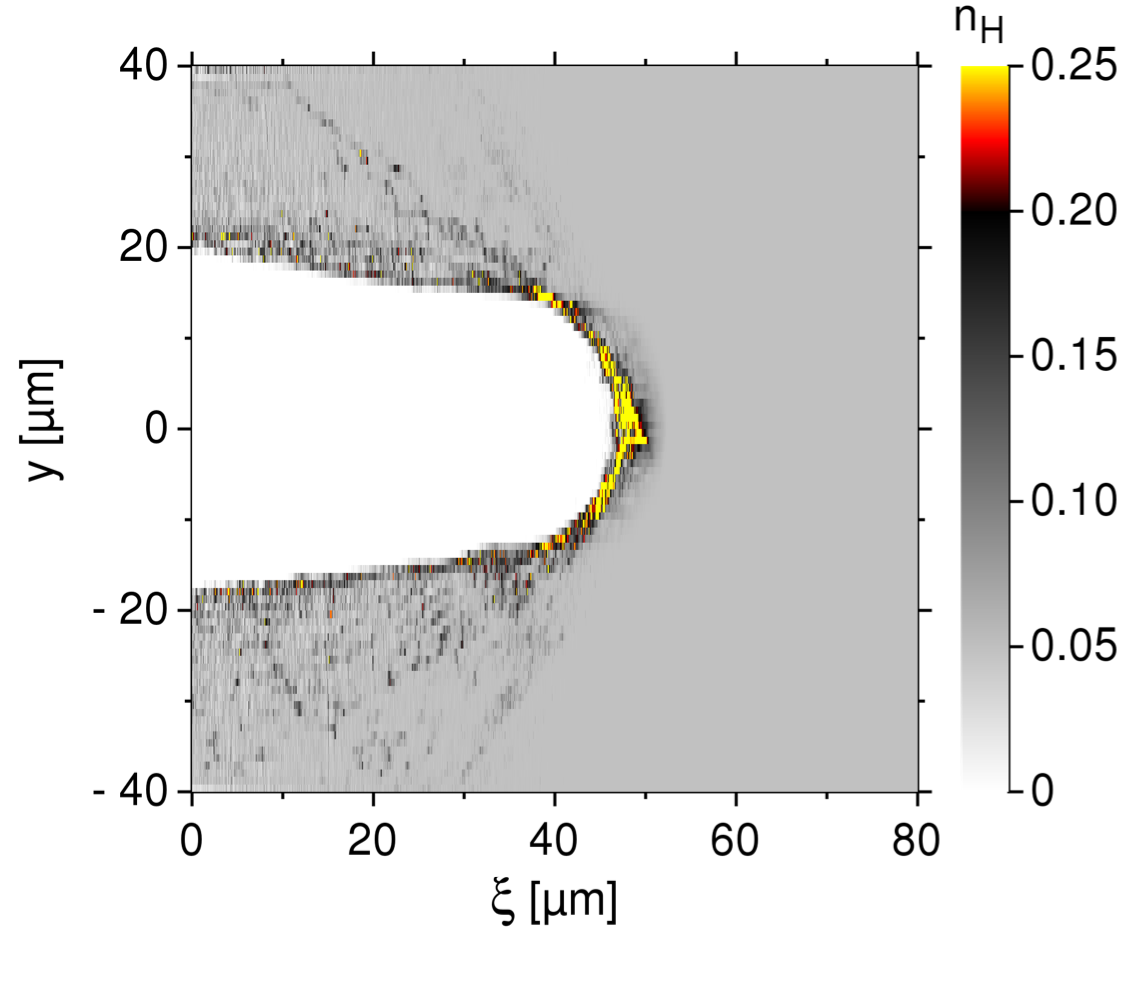}}
\hfill
\subfloat[]{\label{t2_HCl_phase}\includegraphics[width=0.24\textwidth]{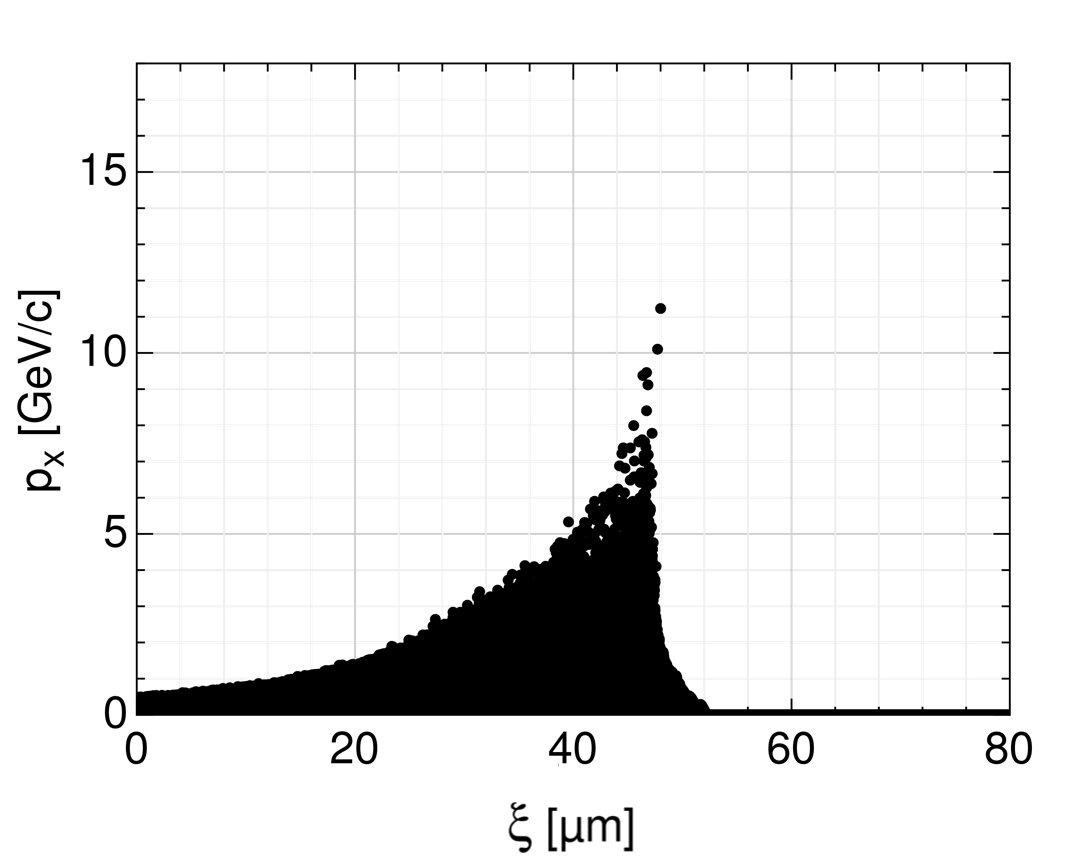}}
\hfill
\subfloat[]{\label{t2_HCl_E_his}\includegraphics[width=0.24\textwidth]{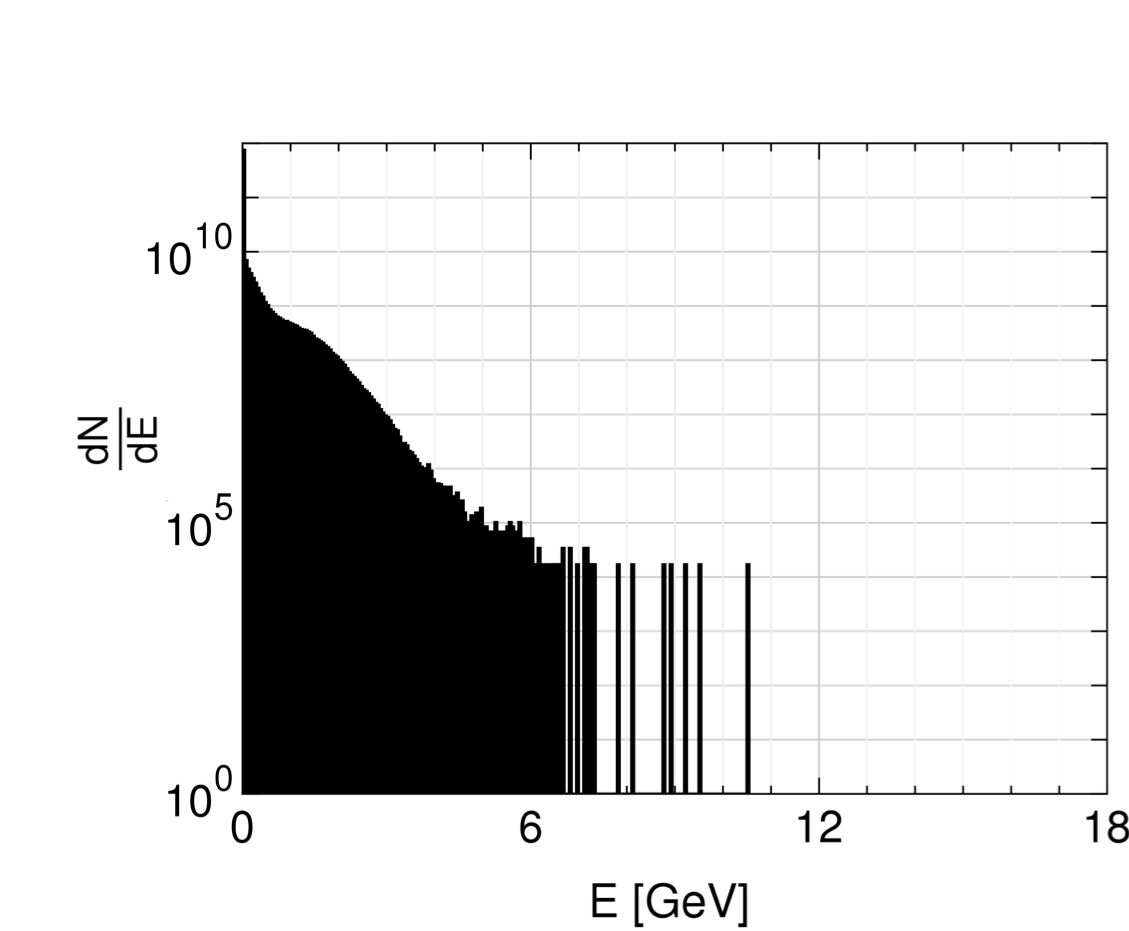}}
\hfill
\subfloat[]{\label{t2_HCl_P_his}\includegraphics[width=0.24\textwidth]{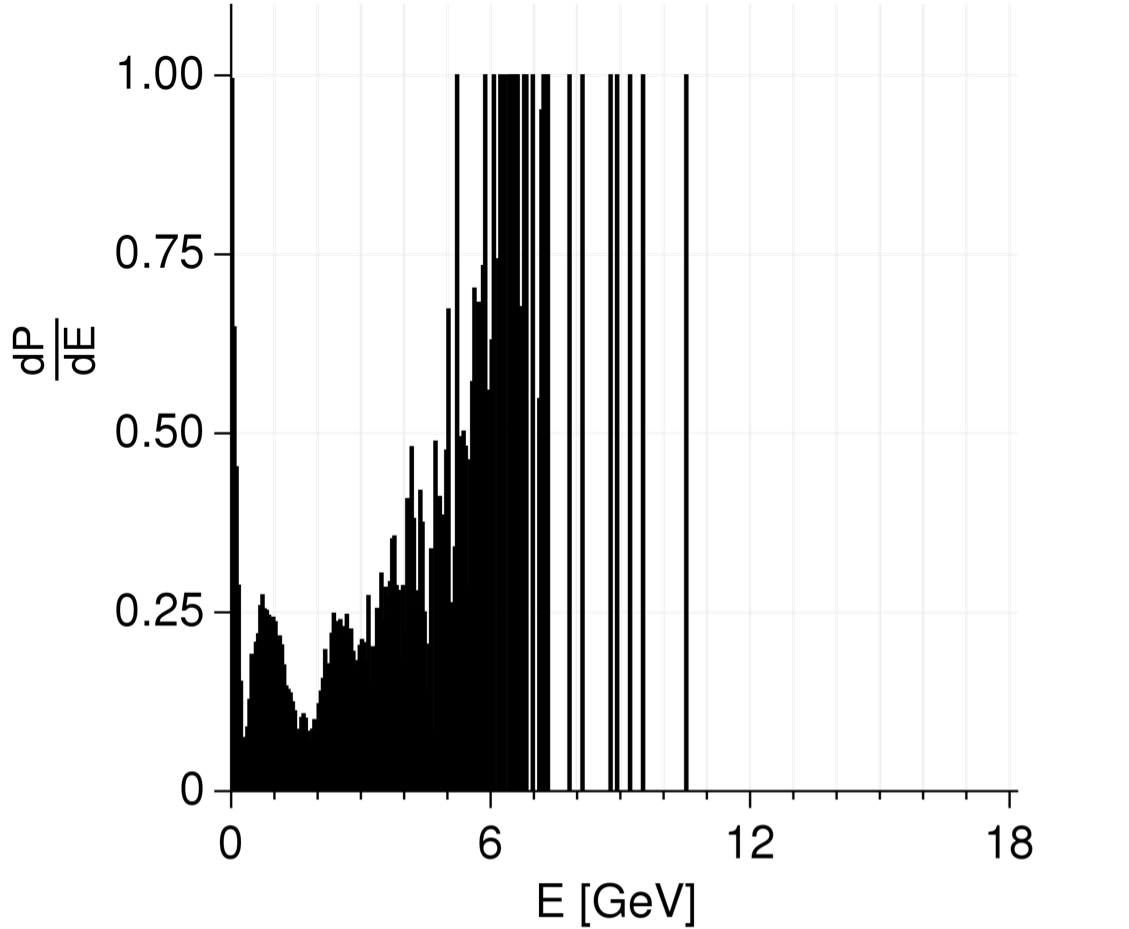}}
\caption{Simulation results for HT (upper plots) and HCl (lower) plasma at the time $t_{\mathrm{2}}=1600$\,fs. Figs.\,(\ref{t2_HT_density}),(\ref{t2_HCl_density}) show the proton densities, (\ref{t2_HT_phase}),(\ref{t2_HCl_phase}) the $p_\mathrm{x}$-$\xi$ the phase-space, (\ref{t2_HT_E_his}),(\ref{t2_HCl_E_his}) the energy histograms and (\ref{t2_HT_P_his}),(\ref{t2_HCl_P_his}) the polarization distributions. The laser peak amplitude is located at $\xi=50.5$\,$\mu$m in the HT and at $\xi=50$\,$\mu$m in the HCl simulation.}
\label{Zeit2}
\end{figure}

\begin{figure}[t]
\centering
\subfloat[]{\label{t3_HT_density}\includegraphics[width=0.24\textwidth]{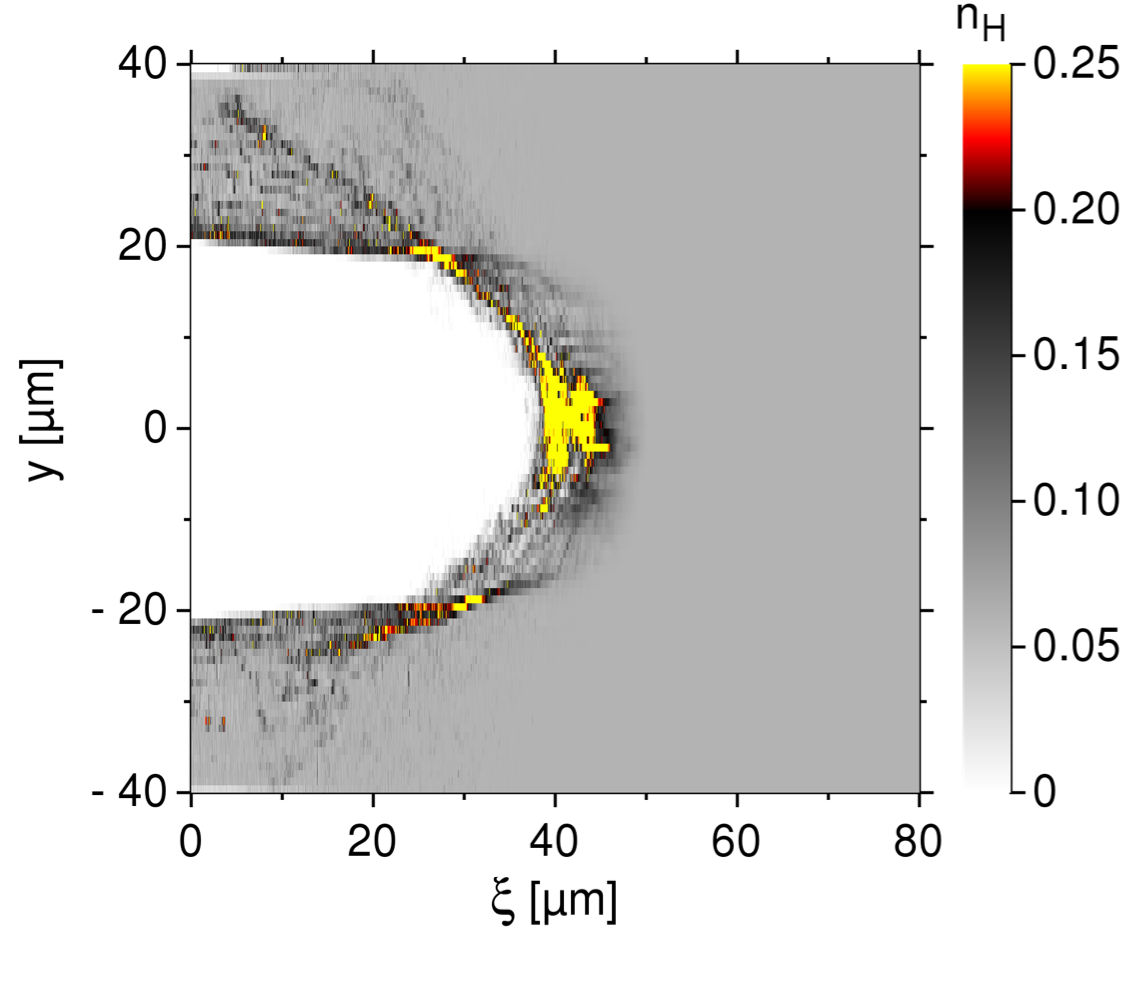}}
\hfill
\subfloat[]{\label{t3_HT_phase}\includegraphics[width=0.24\textwidth]{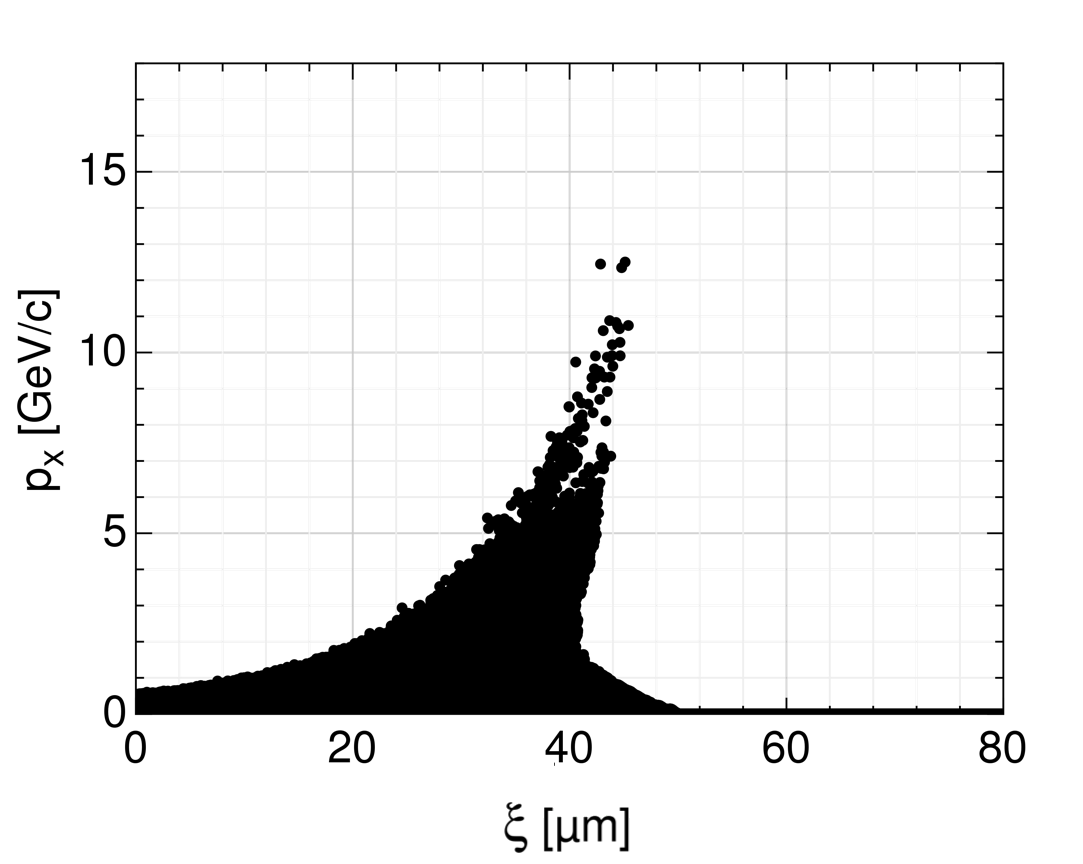}}
\hfill
\subfloat[]{\label{t3_HT_E_his}\includegraphics[width=0.24\textwidth]{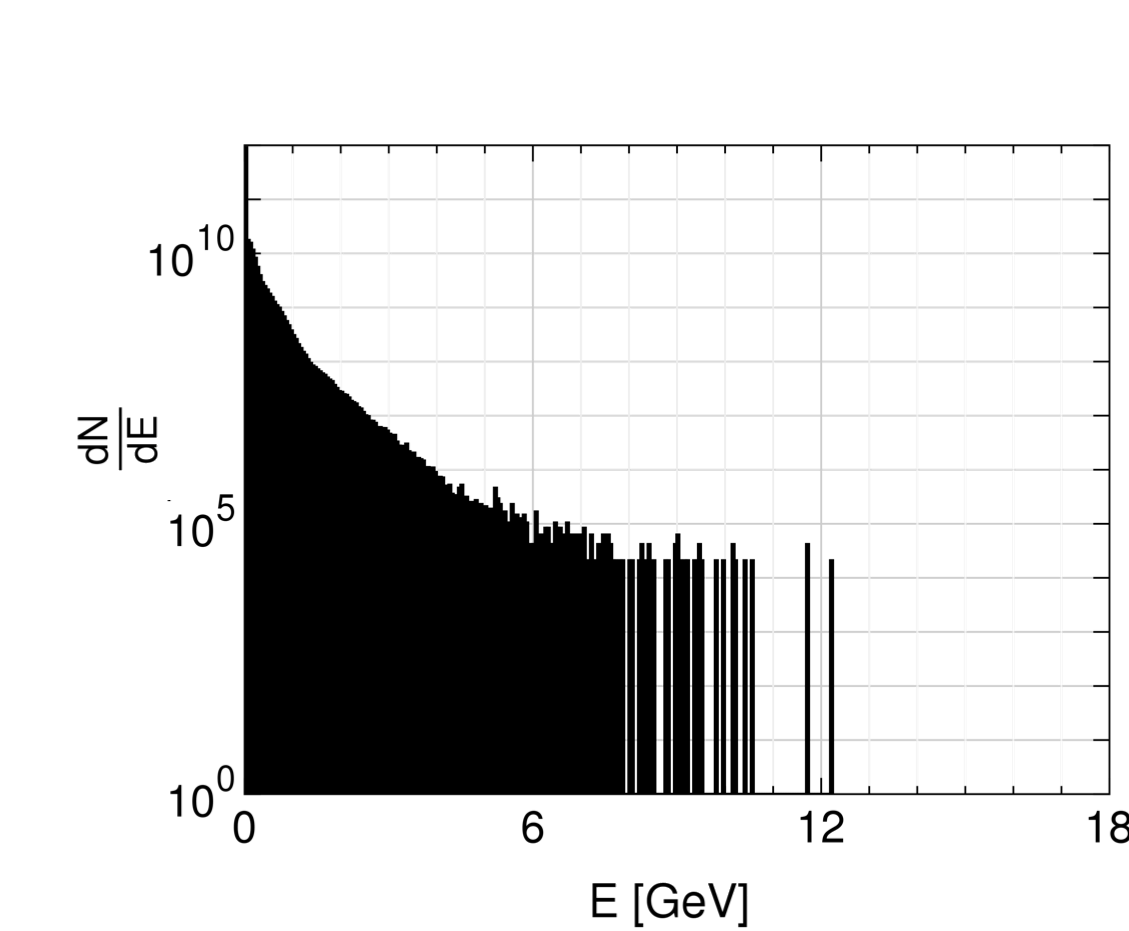}}
\hfill
\subfloat[]{\label{t3_HT_P_his}\includegraphics[width=0.24\textwidth]{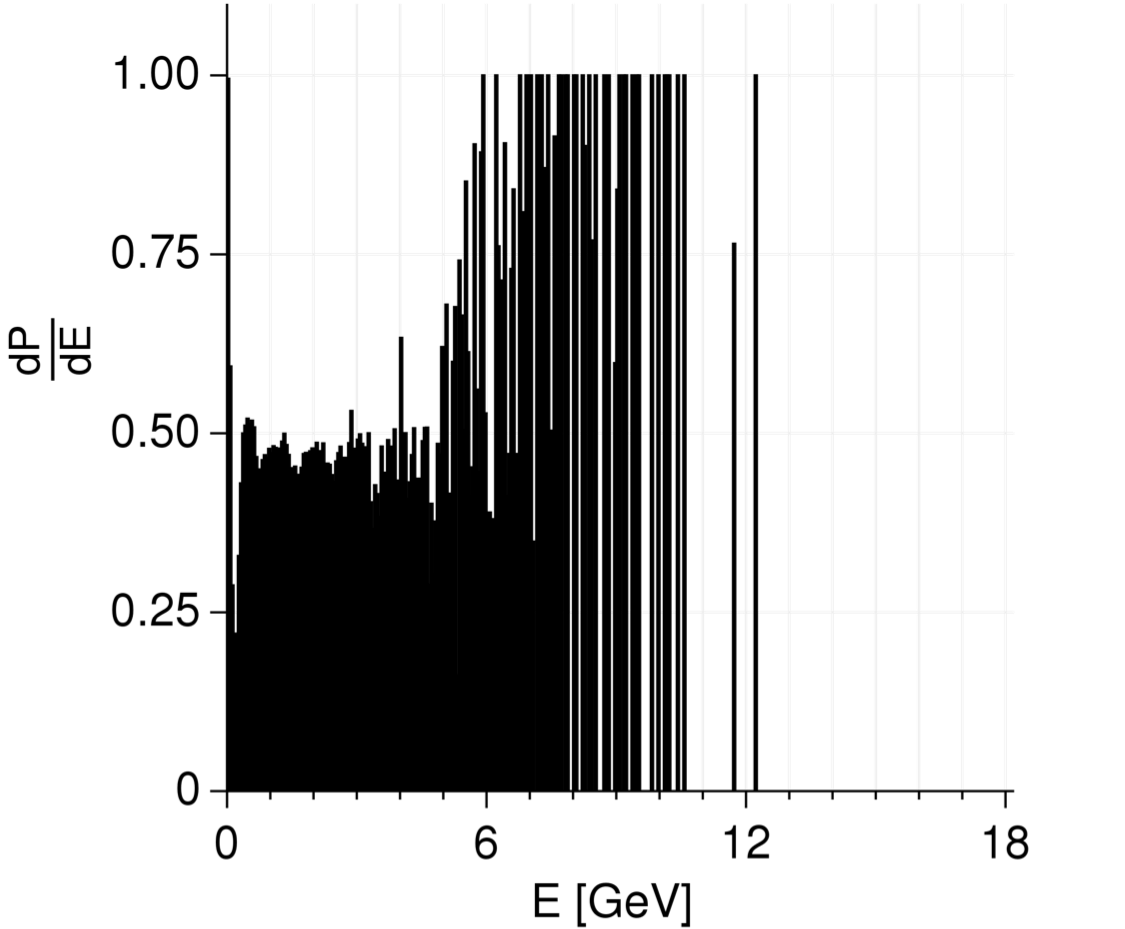}}\\
\subfloat[]{\label{t3_HCl_density}\includegraphics[width=0.24\textwidth]{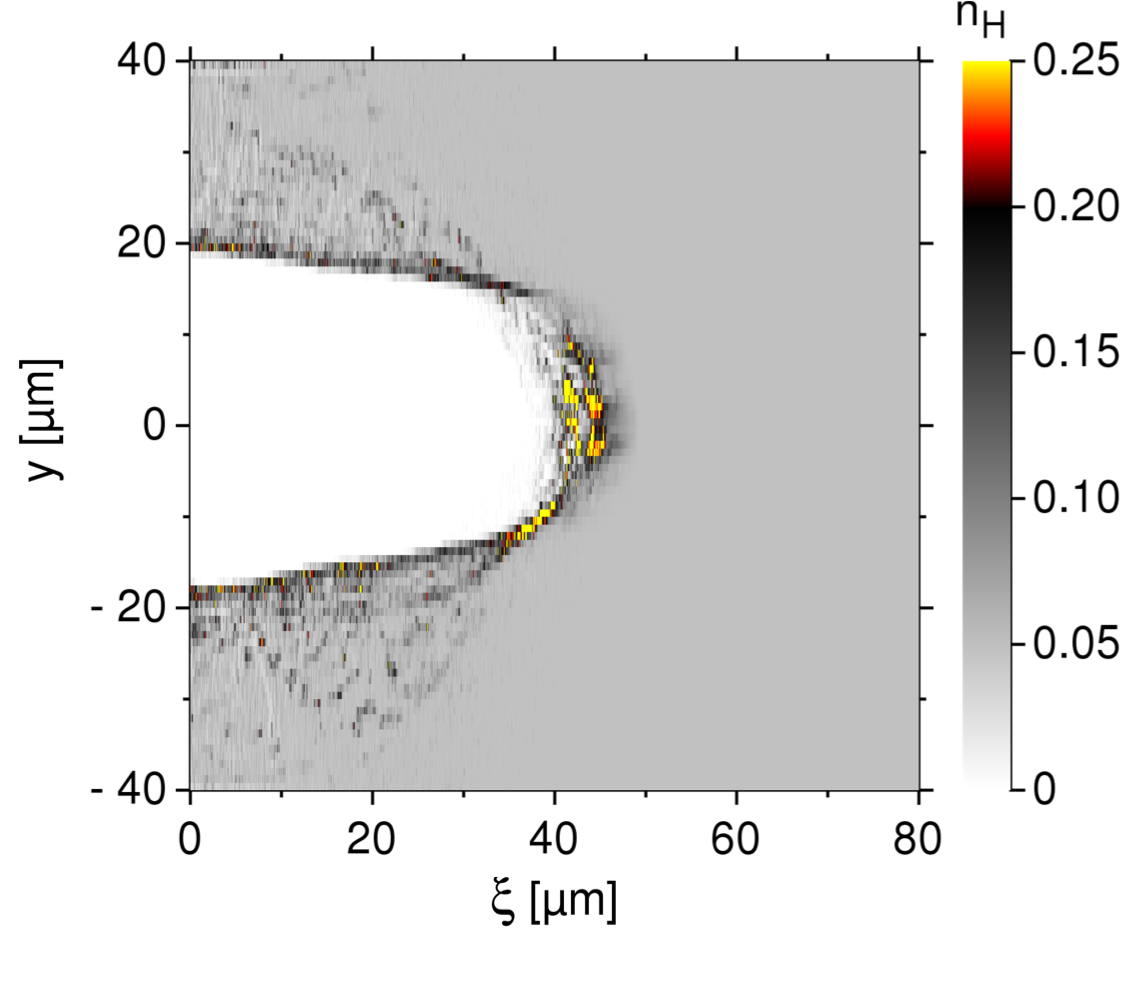}}
\hfill
\subfloat[]{\label{t3_HCl_phase}\includegraphics[width=0.24\textwidth]{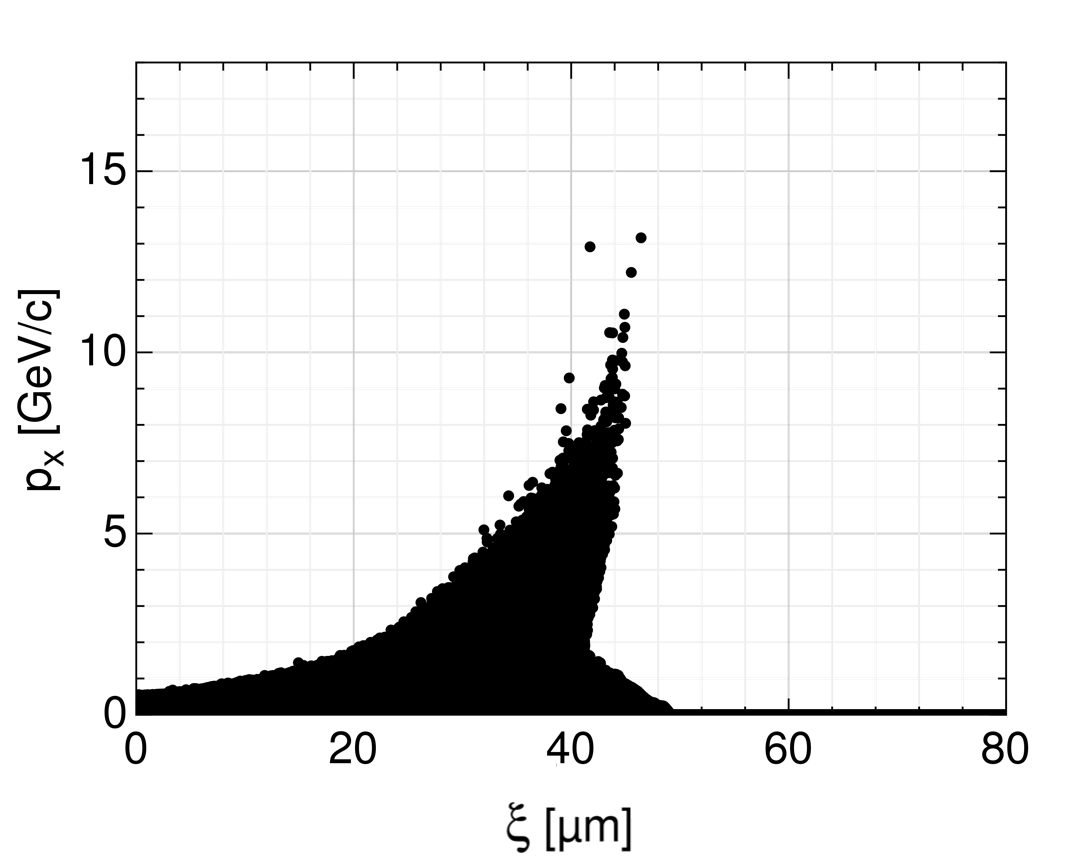}}
\hfill
\subfloat[]{\label{t3_HCl_E_his}\includegraphics[width=0.24\textwidth]{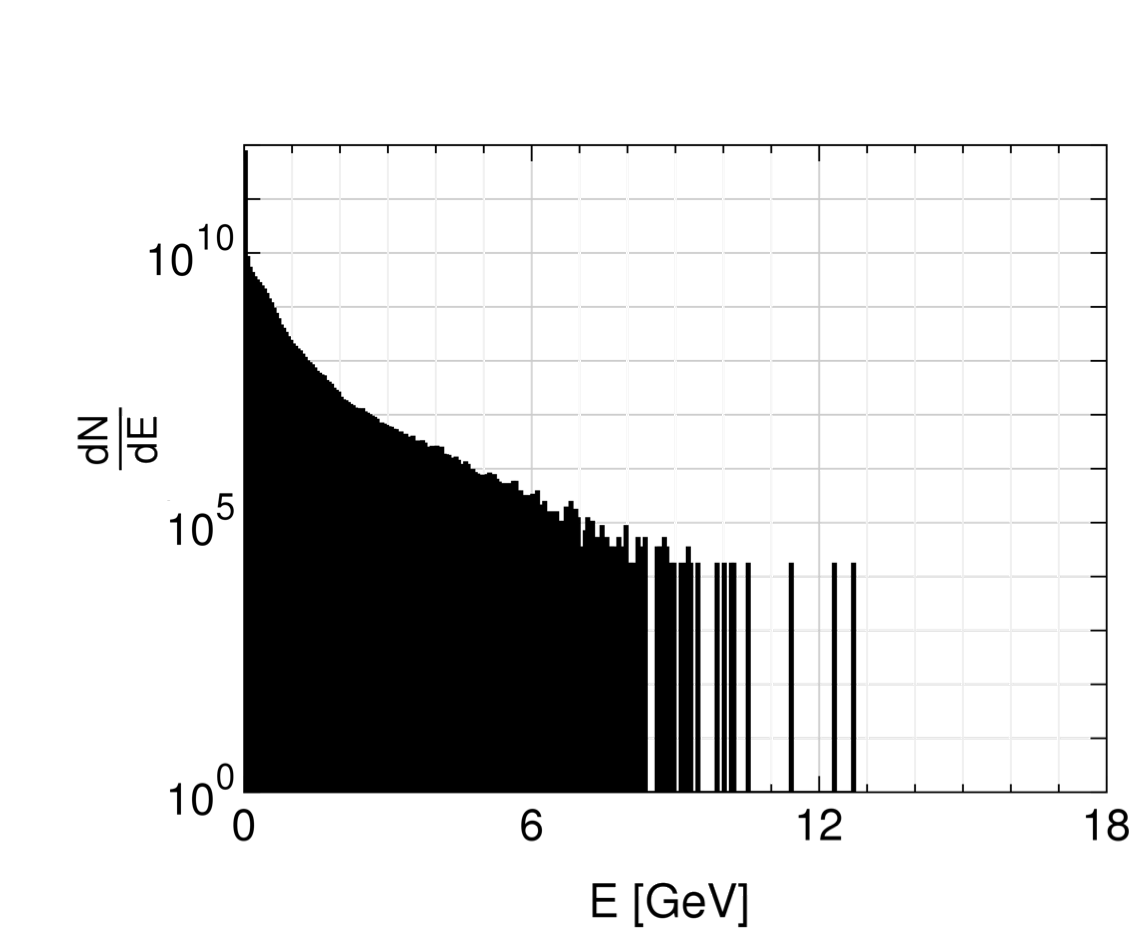}}
\hfill
\subfloat[]{\label{t3_HCl_P_his}\includegraphics[width=0.24\textwidth]{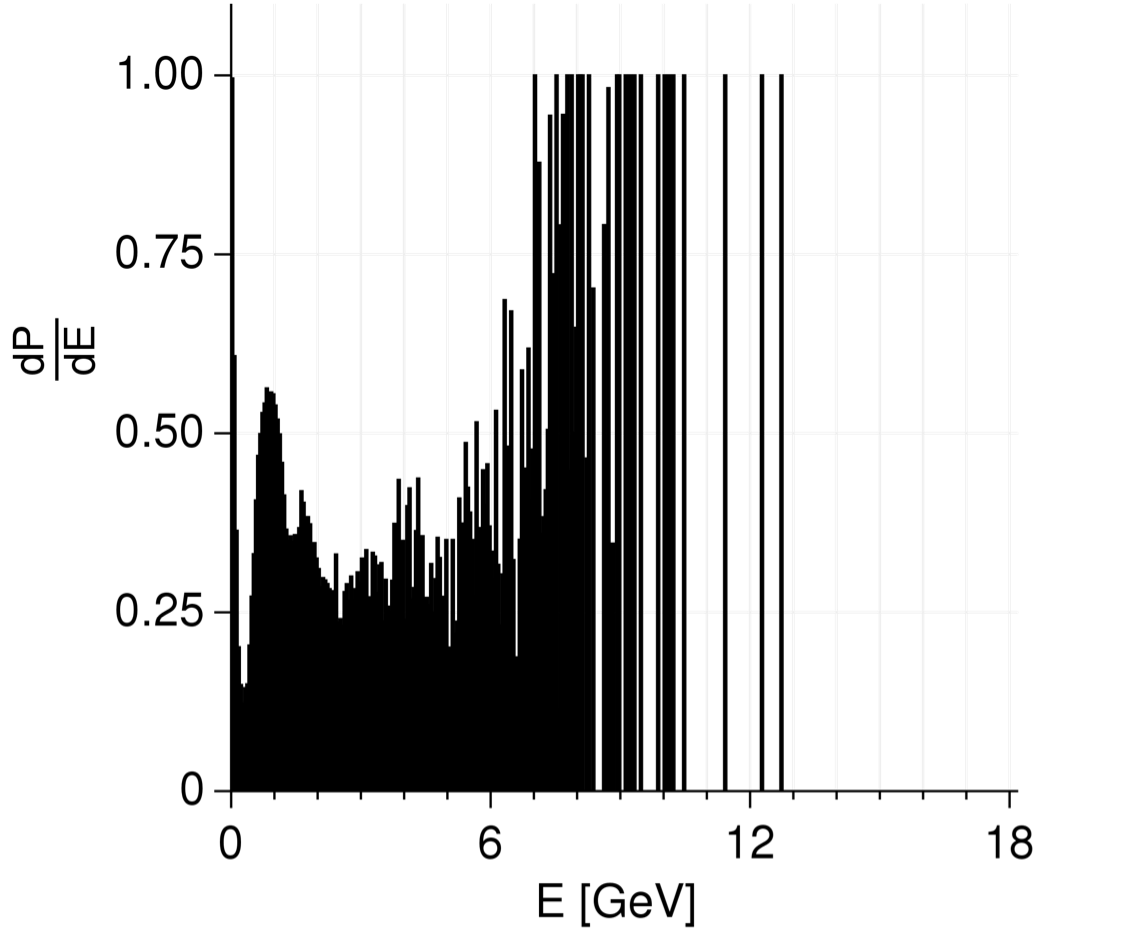}}
\caption{Simulation results for HT (upper plots) and HCl (lower) plasma at the time $t_{\mathrm{3}}=2000$\,fs. Figs.\,(\ref{t3_HT_density}),(\ref{t3_HCl_density}) show the proton densities, (\ref{t3_HT_phase}),(\ref{t3_HCl_phase}) $p_\mathrm{x}$-$\xi$ the phase-space, (\ref{t3_HT_E_his}),(\ref{t3_HCl_E_his}) the energy histograms and (\ref{t3_HT_P_his}),(\ref{t3_HCl_P_his}) the polarization distributions. The laser peak amplitude is located at $\xi=49.5$\,$\mu$m in the HT and at $\xi=48$\,$\mu$m in the HCl simulation.}
\label{Zeit3}
\end{figure}

In the subsequent three figures the proton densities, the $p_{\mathrm{x}}$-$\xi$ phase space for the $x$-component of the kinetic momentum \textbf{p}, the energy spectra and the polarization distributions are represented at three different times $t_{\mathrm{1}}=800$\,fs (Fig.\,\ref{Zeit1}), $t_{\mathrm{2}}=1600$\,fs  (Fig.\,\ref{Zeit2}) and $t_{\mathrm{3}}=2000$\,fs  (Fig.\,\ref{Zeit3}), comparing hydrogen-tritium (HT) and hydrogen-chloride (HCl) in each figure. These times correspond to the  propagation distances of 240\,$\mu$m, 480\,$\mu$m and 600\,$\mu$m of the moving frame.

The proton densities at time $t_{\mathrm{1}}$ in Figs.\,(\ref{t1_HT_density}) and (\ref{t1_HCl_density}) show a channel with a diameter larger than $40$\,$\mu$m shortly after its formation. After a further stabilization phase, the channel structures at time $t_{\mathrm{2}}$ in Figs.\,(\ref{t2_HT_density}) and (\ref{t2_HCl_density}) look very similar. The only difference is a slightly larger proton density in the HT plasma in front of the channel. The accumulation of protons and the vanishing channel structure at time $t_{\mathrm{3}}$ in Figs.\,(\ref{t3_HT_density}) and (\ref{t3_HCl_density}) shows that the laser has started depleting in both plasmas. At this advanced time the hydrogen density in the front of the channel structure in HT is visibly enhanced compared to the HCl simulation case.

The phase-space plots with a peak of height $2.5$\,GeV/c at $\xi=50$\,$\mu$m are nearly identical at time $t_{\mathrm{1}}$ in Figs.\,(\ref{t1_HT_phase}) and (\ref{t1_HCl_phase}). A comparison to the proton densities in Figs.\,(\ref{t1_HT_density}) and (\ref{t1_HCl_density}) shows that the dense (yellow) spots at the channel walls are rather cold ($E\leq1$\,GeV) protons which are not further accelerated. The phase-space distributions at time $t_{\mathrm{2}}$ in Figs.\,(\ref{t2_HT_phase}) and (\ref{t2_HCl_phase}) are similar in shape and height in both cases but the HCl simulation indicates a slightly higher proton density around $6$\,GeV/c. An increased peak level of more than $10$\,GeV/c in both diagrams demonstrates a mean acceleration force in the range of $30$\,TeV/m. A comparison of the positions of those protons with large momentum with the density plots confirms that this field strength is reached in front of the channel. The phase-space plot at time $t_{\mathrm{3}}$ for HT shows similar characteristics like the one for HCl. In both simulations the highest momenta are reached shortly before the laser depletes (cf. Fig.\,\ref{Zeit3}). At this time the peak density in Figs.\,(\ref{t3_HT_phase}) and (\ref{t3_HCl_phase}) is $14$\,GeV/c but only a small fraction of protons gain a momentum higher than $10$\,GeV/c. The phase-space density between $2.5$\,GeV/c and $10$\,GeV/c is much larger than at the previous times, nevertheless most protons still have a momentum smaller than $2.5$\,GeV/c. Though the maximal momentum for HCl is a bit higher, the corresponding protons with large momentum are represented by single PIC particles only. This artifact appears due to the strongly limited number of particles per cell.

Similar to the observations in the density and the phase-space plots, only a few differences are visible in the energy histograms and the polarization distributions at $t_{\mathrm{1}}$ in Fig.\,\ref{Zeit1}. In Figs.\,(\ref{t2_HT_E_his}) and  (\ref{t2_HCl_E_his}) d$N$/d$E$ linearly decreases for HT plasma, while the HCl histogram shows a higher particle number at lower ($E\approx1$\,GeV) and higher ($E>5$\,GeV) energies. The single bars both in the HT and the HCl polarization spectrum in Figs.\,(\ref{t2_HT_P_his}) and (\ref{t2_HCl_P_his}) result from single PIC particles at highest energies (cf. artifact in the phase space plots). For lower energies around $5$\,GeV the polarization distribution for HT indicates that at least 50\% of the initial polarization can be preserved. In contrast to the distribution in Fig.\,(\ref{t2_HCl_P_his}) that demonstrates that protons at even higher energies do not loose more than 30\% of their polarization if they are accelerated in HCl plasma. In the energy histograms at time $t_{\mathrm{3}}$ in Figs.\,(\ref{t3_HT_E_his}) and (\ref{t3_HCl_E_his}) the bars drop equally fast between $0$\,GeV and below $2$\,GeV. Afterwards, the spectra show small difference as for HT the decrease in the range between $2$\,GeV and $5$\,GeV is still linearly but with a smaller gradient, whereas the histogram for HCl shows a significantly shallower slope. In both cases, d$N$/d$E$ is almost constant within the $5$\,GeV to $8$\,GeV range. For higher energies ($E>8$\,GeV) there are only a few PIC particles present. The maximum energy is slightly higher for HCl, which confirms the observations for the corresponding phase-space plots. The HT polarization distribution shows small amplitude fluctuations around d$P$/d$E=0.5$ for energies up to $6$\,GeV (cf. Fig\,(\ref{t3_HT_P_his})). In this energy range the HCl distribution is less stable and drops to much smaller values, while both spectra exhibit strong noise in the high-energy range well above $6$\,GeV.

\newpage
\section{Conclusion}

In three-dimensional PIC simulations we demonstrate the acceleration of pre-polarized protons in HT and HCl plasma over a distance of 600\,$\mu$m. We show that the proton energies easily reach the $10$\,GeV level, while the polarization is largely conserved for at least those protons with energy above $5$\,GeV. To analyse the polarization of proton with higher energies in detail a larger number of PIC particles per cell is needed in further simulation studies. Polarized HCl gas seems to be the better target choice compared to HT gas because we observe slightly higher energies and momenta. Besides this advantage, a pre-polarized HCl target is already under construction at Forschungszentrum J\"ulich and  first measurements  of high-energetic polarized proton beams are in preparation at the multi-PW laser facility SULF  (Shanghai).

\ack
This work has been carried out in the framework of the \textit{Ju}SPARC (J\"ulich Short-Pulse Particle and Radiation Center) project and has been supported by the ATHENA consortium. We further acknowledge the computing resources on grant VSR-JPGI61 on the supercomputer JURECA \cite{Jureca}. The Chinese authors acknowledge support through the Strategic Priority Research Program of Chinese Academy of Sciences (Grant No. XDB 16010000), the National Science Foundation of China (No. 11875307) and the Recruitment Program for Young Professionals.

\end{document}